\documentclass[aps,pra,twocolumn,nopacs,floatfix]{revtex4}
\usepackage{ulem}
\usepackage{epsfig}
\usepackage{graphicx}
\usepackage{dcolumn}
\usepackage{amsthm,amsmath}
\usepackage{comment}
\usepackage[colorlinks]{hyperref}
\usepackage{xcolor}
\usepackage{mathrsfs}
\usepackage{longtable}
\usepackage{amsmath}

\begin{document}

\title{Dispersion $C_3$ coefficients for physisorption of heavy ions and atoms with graphene and carbon nanotubes}

\author{$^1$Harpreet Kaur}
\author{$^2$Neelam Shukla}
\author{$^2$Rajesh Srivastava}
\author{$^1$Bindiya Arora}
\email{bindiya.phy@gndu.ac.in}

\affiliation{$^1$Department of Physics, Guru Nanak Dev University, Amritsar, Punjab 143005, India}
\affiliation{$^2$Department of Physics, Indian Institute of Technology Roorkee, Roorkee, Uttarakhand 247667, India}

\begin{abstract}
In the present work, the calculations have been performed for dispersion interaction between heavy elements (Zn$^+$, Cd$^+$, Hg$^+$, Pb$^+$, Zn, Cd, Hg, and Pb) with graphene and carbon nanotubes by evaluating van der Waals $C_3$ coefficients using well-known Lifshitz theory. The dispersion coefficients are expressed in terms of reflection coefficients of graphene and carbon nanotubes which are  calculated within the framework of Dirac model. In addition, accurate values of dynamic dipole polarizabilities at imaginary frequencies of considered ions and atoms which are vital, have been calculated using relativistic methods. The comparisons of our calculated static dipole polarizabilities for the considered elements with the values reported in the literature are also presented. The dispersion interactions of the considered heavy elements with the graphene and carbon nanotubes of different radii in a wide range of separation distances have been studied. The results have also been analysed with another subtle variable, i.e., gap parameter of graphene wall. The interaction coefficients obtained for both the materials, i.e., for graphene and carbon nanotubes are mutually compared and it is found that graphene can be said as preferable material for adsorption of these toxic heavy elements.

\end{abstract}

\maketitle

\section{Introduction}
Metals occur naturally in the earth's crust, and their contents in the environment influence the ecologies in many habitats~\cite{morais2012heavy}. Among various metals, heavy atoms and their ions with high atomic weights and large densities are found to be toxic to the human body even when present in trace amounts in various environmental matrices~\cite{jaishankar2014toxicity}. This has lead to growing public health concerns about heavy metal pollution. Non-biodegradable characteristics of these elements have the capability of causing detrimental effects to the entire biodiversity~\cite{tovar2018heavy, nagajyoti2010heavy}. The high solubility of heavy ions leads to contamination of natural resources such as water and soil, which as a consequence gets accumulated in organisms and enters the food chain leading to a process of biomagnification~\cite{ali2019trophic}. Excessive exposure of Zn can lead to brain, respiratory and gastrointestinal syndrome~\cite{plum2010essential}. Cd species can cause skeletal damage as a secondary response to kidney damage or direct action on the bone cells, whereas Hg species being carcinogenic cause adverse effects on the development of human brain~\cite{mahurpawar2015effects}. Hematopoietic, renal, reproductive, and central nervous systems are vulnerable towards the dangers caused by exposure to the high level of Pb species~\cite{assi2016detrimental}. {The primary sources of these elements are various industrial activities, natural resources, agriculture, and untreated disposal of domestic waste~\cite{nagajyoti2010heavy, wuana2011heavy}.} Therefore, accurate and accessible detection of these toxic elements is necessary to ensure  environmental quality control and early warning capabilities to avoid public safety adversity.

Detection of these elements with various conventional materials like clay, its minerals, zeolites, activated carbon, fullerenes, biomaterials, etc., has been done previously~\cite{uddin2017review, hong2019heavy, burakov2018adsorption}. Further, nanomaterials show great technological advances in a wide range of applications due to extraordinary properties as compared to their bulk counterparts~\cite{poole2003introduction}. The rapid growth of nanomaterials for various applications has seen a boost after the discovery of graphene. Many breakthroughs in the research of graphene have been observed in the last decade due to its large surface-to-volume ratio, thin structure, and interface interactions. Graphene and graphene-based nanostructures render unique mechanical, electrical, optical, and thermal properties~\cite{papageorgiou2017mechanical, phiri2018comparative, fan2014thermal, falkovsky2008optical} that  have significantly made this material as one of the most studied two-dimensional (2D) material in condensed matter physics contributing in various applications like electrochemical devices, solar cells, plasmonic, purifiers, sensors etc.~\cite{kavan2013application, nguyen2016promising, grigorenko2012graphene, dervin20162d, mao2014nanocarbon}. Besides this, one dimensional (1D) allotrope of carbon; single-walled carbon nanotubes (SWCNT) with diameter less than 50 nanometers (nm) having different configurations exhibit similar properties as that of single-layer graphene~\cite{torres2017mesoscale}. 

It has been observed both experimentally and theoretically that adsorption technology can monitor trace amounts of heavy metals. Chemical adsorption of adsorbate on a graphene-based system can modify its properties, providing a non-reversible binding of the atom or molecule to the surface. Therefore, physical adsorption is always preferred due to its reversible nature.  Graphene and carbon nanotubes (CNT) have been extensively explored for physical adsorption of some of the heavy ions, dye molecules, and hydrogen molecules~\cite{shtepliuk2017interaction, yusuf2015applications, niemann2008nanomaterials, henwood2007ab} for sensor applications.  Even now, the interaction studies for physisorption of heavy elements with graphene and CNT have been done theoretically and calculations are performed using Density functional theory (DFT)~\cite{ou2015physisorption, mashhadzadeh2018dft, petrushenko2019hydrogen, lazic2005role, silvestrelli2012adsorption}. Abdesalam \textit{et al.}~\cite{abdelsalam2019first} and Shtepliuk \textit{et al.}~\cite{shtepliuk2017interaction} studied the adsorption of toxic heavy elements on graphene-based system. However, a study by Oyetade \textit{et al.}~\cite{oyetade2017experimental} showed nitrogen-functionalized carbon nanotubes as a good reusable adsorbent for the removal of Pb$^{+2}$ and Zn$^{+2}$ from wastewater. 

 Other studies for physical adsorption of microparticles with the material given by generalized {Lifshitz} theory have been conducted using \textit{ab initio} calculations~\cite{jiang1984dispersion, rauber1982substrate, zaremba1976van}. The theory explains the interactions of atoms or molecules with material walls in both retarded and non retarded regimes giving rise to Casimir-Polder and van der Waals (vdW) forces~\cite{klimchitskaya2020casimir, bordag2006lifshitz, bordag2009advances}. These forces find diverse applications in circuit technology, adsorption, quantum reflections, and Bose condensation~\cite{bordag2009advances, lin2004impact, bezerra2008lifshitz, zaremba1976van, tao2014physical}.  {Lifshitz} theory gave a generalization of both these interaction forces in which the strength of the attractive forces is expressed in terms of dispersion $C_3$ coefficient~\cite{caride2005dependences}. Dispersion coefficients have been calculated for a number of material walls, including metals, semiconductors, insulators, and dielectrics, by taking the optical properties into account~\cite{caride2005dependences, arora2014van, derevianko1999high, lach2010noble, dutt2020van, blagov2007van, kaur2016dispersion}. These dispersion coefficients were also measured experimentally using atomic force microscopy (AFM) and spectroscopy techniques~\cite{fichet2007exploring, lepoutre2009dispersive, lonij2011can, schneeweiss2012dispersion}. Such studies were reported for applications in hydrogen sensing, storage and designing an up-gradation technology for batteries~\cite{blagov2005van, blagov2007van, bordag2006lifshitz}. 
  
  In the present work,  {we particularly focus on} the interaction of heavy elements with carbon-based systems - graphene and CNT which are considered as two-dimensional free-electron gas. Reflection coefficients of these materials are important contributors  to the calculation of the dispersion coefficients.  {Out of the models proposed in the literature for the evaluation of reflection coefficients, the Dirac model approach is preferred due to its providing results in close agreement with experiment~\cite{klimchitskaya2015comparison}.} Previously, studies conducted were based on this approach for the interaction of alkali atoms, alkaline ions, noble gas molecules, hydrogen atom, and hydrogen molecule with graphene and CNT wall~\cite{kaur2015dispersion, bordag2006lifshitz}. Accurate values of the polarizability of microparticles at imaginary frequencies are necessary to compute $C_3$ coefficients between the microparticle and the material wall given by generalized Lifshitz theory. In this paper, we have calculated the $C_3$ dispersion coefficients for interaction of microparticle with graphene and CNT wall along with the evaluation of static and dynamic polarizabilities of heavy ions and atoms at imaginary frequencies using the sum-over-states approach. There are a few studies that have reported only static polarizabilities. Most of these have used non-relativistic methods but for such heavy elements it is necessary to adopt a relativistic approach as we have done in the present work for the reliable calculations of atomic properties. 
  
   The outline of the paper is as follows. In Sec~\ref{Sec II}, we give a brief overview of the theory. Sec.~\ref{Sec III} contains the evaluated values of static dipole polarizability of heavy ions and atoms. The dynamic dipole polarizabilities for ions and atoms are also presented in the same section. In addition to this, the dispersion coefficients between considered ions or atoms and materials have been discussed. We have also compared the results of dispersion coefficients for graphene and CNT. The dependency of the gap parameter on interaction coefficients is also discussed in this section. Atomic units (a.u.) have been used throughout the paper unless stated otherwise.
  
\section{Theory}\label{Sec II}
\subsection{Dispersion coefficient}
Generalized Lifshitz formula for the non-retarded vdW interaction energy of atoms or molecules with graphene and CNT wall using proximity force approximation (PFA) can be written in terms of dispersion coefficients $C_3$ for a separation distance $a$ in the following form ~\cite{churkin2011dispersion}
 
\begin{equation}
E(a)= -\frac{C_3 (a)}{a^3}.
\end{equation}
The $C_3$ coefficient due to interaction between graphene and microparticle is expressed in the terms of reflection coefficients $r_{TM}$ and $r_{TE}$ as follows ~\cite{churkin2011dispersion, arora2014coefficients}

\begin{eqnarray}
& & C_3(a) = \frac{1}{16\pi}\int_{0}^\infty\alpha(\iota\xi) d\xi \int_{2a\xi\alpha_{fs}}^\infty y^2e^{-y}dy 
\nonumber \\
& & \times \left[2r_{TM} - (r_{TM} + r_{TE})\frac{4a^2\xi^2\alpha_{fs}^2}{y^2} \right],
									\nonumber\\
									& &
\label{Eq.2}
\end{eqnarray}
whereas this coefficient for CNT of radius $R$ becomes radius dependent and can be expressed as:
\begin{eqnarray}\label{Eq.3}
& & C_3(a,R) = \frac{1}{16\pi}\sqrt{\frac{R}{R+a}}\int_{0}^\infty\alpha(\iota\xi) d\xi \int_{2a\xi\alpha_{fs}}^\infty ye^{-y} dy 
\nonumber \\ 
& & \times \left(y-\frac{a}{2(R+a)}\right) 
\times \left[2r_{TM} - (r_{TM} + r_{TE})\frac{4a^2\xi^2\alpha_{fs}^2}{y^2}\right]. 
									\nonumber \\
									& &
\end{eqnarray}
In both the above expressions, $\alpha$ is the dynamic dipole polarizability of the ion or atom over imaginary frequencies $\iota\xi$ and $\alpha_{fs}$ is the fine structure constant~\citep{churkin2011dispersion}. $y$ is a dimensionless variable given by $y=2aq$, where $a$ is the separation distance and $q=\sqrt{k^2 + \xi^2}$, dependent on wave vector $k$~\cite{churkin2011dispersion}. For the evaluation of these reflection coefficients, two models have been proposed in the literature for graphene and CNT. These two are Dirac~\cite{geim2009graphene, bordag2009casimir, neto2009electronic} and hydrodynamic models~\cite{bordag2001new, bordag2006lifshitz}. In hydrodynamic model, graphene is taken as infinitesimally thin positively charged sheet with a continuous fluid of mass and negative charge densities. The dispersion relation for quasiparticles in graphene is quadratic with respect to the momentum. However, this model does not take into account some properties of graphene which are important at low energies and due to this reason it overestimates the vdW interactions. In the Dirac model, the quasiparticles in graphene are considered to be Dirac Fermions moving with Fermi velocity and follow linear dispersion law. This model has provided results in accord with experimental values~\cite{klimchitskaya2015comparison}. In this work,  Dirac model has been implemented for determination of dispersion coefficients. Under this framework, the explicit forms of two components of the reflection coefficients are given by~\cite{bordag2009casimir}

\begin{equation}\label{Eq.4}
r_{TM}= \frac{\alpha_{fs} q \phi(\tilde{q})} {2 \tilde{q}^2 + \alpha_{fs} q \phi(\tilde{q})},
\end{equation}
\begin{equation}\label{Eq.5}
r_{TE}= \frac{\alpha_{fs} q \phi(\tilde{q})} {2 \tilde{q} + \alpha_{fs} q \phi(\tilde{q})},
\end{equation}
where $q = \sqrt{k^2 +\xi^2/c^2}$, $\tilde{q}$ is the function of Fermi velocity $v_f$ of massless Fermions and $\phi$ is the polarization tensor. The expressions of these two parameters can be given as~\cite{bordag2009casimir}

\begin{equation}\label{Eq.6} 
\tilde{q}=\sqrt{\frac{\alpha_{fs}^2v_f^2y^2}{4a^2}+(1-\alpha_{fs}^2v_f^2)\alpha_{fs}^2\xi^2},
\end{equation}

\begin{equation}\label{Eq.7}
\phi(\tilde{q})=4\left(\alpha_{fs}\Delta+\frac{\tilde{q}^2-4\alpha_{fs}^2\Delta^2}{2\tilde{q}}\arctan\left(\frac{\tilde{q}}{2\alpha_{fs}\Delta}\right)\right),
\end{equation}
where $\Delta$ is the gap parameter~\cite{neto2009electronic} whose value lies in the range $0 < \Delta < 0.1$ eV. Since the value of gap parameter is still not known, we have taken its value initially as 0.01 throughout the paper unless stated otherwise.

\subsection{Dipole polarizability} 
	The dipole polarizability for an atomic system in the ground state $n$ with a closed core and valence electron(s) can be evaluated by calculating two components of the polarizability as follows~\cite{arora2012multipolar}
	\begin{equation}\label{Eq.8}
	\alpha(\iota\omega)= \alpha_{val}(\iota\omega) + \alpha_c(\iota\omega),
	\end{equation}	 
	where subscript $val$ and $c$ refer respectively to the polarizability contributions due to valence and core orbitals respectively. The dominant contribution to the polarizability is from the valence part which can be further expressed in terms of  Main and Tail contributions. The Main term of $\alpha_{val}$ contains the contributions due to the low lying allowed transitions from the ground state whereas Tail term has the contributions of the transition from  ground to higher states. 
	The Main term of valence contribution can be estimated as follows
	\begin{eqnarray}\label{Eq.9}
& & \alpha_{val}^{Main}(\iota\omega)=\frac{2}{3(2J_n+1)}
\nonumber \\
& & \times \sum_{m > N_c,m\neq n}^I \frac{(E_m-E_n)|\langle\psi_n||\textbf{D}||\psi_m\rangle|^2}{(E_m-E_n)^2+\omega^2}.
                             \nonumber \\
                             & &
	 \end{eqnarray}
In the above equation, $J_n$ is the total angular momentum quantum number of the ground state of the considered atom/ion sum is restricted by including sum over intermediate $m$ states after $N_c$ and up to $I$, where $N_c$ represents the core orbitals and $I$ refers to the bound states up to which we have determined the reduced matrix elements $\langle\psi_n||D||\psi_m\rangle$ in our calculations. We use relativistic all order method and multiconfigurational Dirac-Fock (MCDF) approximation for ions and atoms respectively to compute the matrix elements used for the Main term calculation. In order to do reliable calculations and avoid any uncertainties, we use the experimental excitation energy values $E_i$ of the corresponding states for Main term taken from National Institute of Standards and Technology (NIST) database~\cite{NIST_ASD}.
	
	Similarly, the Tail term is evaluated using the following equation
		\begin{equation}\label{Eq.10}
	 \alpha_{val}^{Tail}(\iota\omega)=\frac{2}{3(2J_n+1)}\sum_{m > I} \frac{(\epsilon_m-\epsilon_n)|\langle\psi_n||{\bf D}||\psi_m\rangle_{DHF}|^2}{(\epsilon_m-\epsilon_n)^2+\omega^2},
	 \end{equation}
where $\langle\psi_n||{\bf D}||\psi_m\rangle_{DHF}$ are the E1 reduced matrix elements obtained using DHF method.  $\textbf{D}$ is the dipole operator defined as $\textbf{D}= -e\sum_j\textbf{r}_j$ with $\textbf{r}_j$ being position of a $j$th electron, and the sum $m>I$ corresponds to the excited states whose matrix elements are not accounted in the Main term. The energies calculated using DHF method are referred by $\epsilon_i$.
The calculations of the core polarizabilities of both ions and atoms are carried out in the DHF method using the following expression, 
	 
	 \begin{equation}\label{Eq.11}
	 \alpha_{c}(\iota\omega)=\frac{2}{3(2J_n+1)}\sum_{a}^{N_c}\sum_{m}^{I} \frac{(\epsilon_m-\epsilon_a)|\langle\psi_a||\textbf{D}||\psi_m\rangle_{DHF}|^2}{(\epsilon_m-\epsilon_a)^2+\omega^2},
	 \end{equation}
where $a$ refers to the core orbitals while $m$ includes valence or empty orbitals. 
The evaluation of the core correlation using the above expression does not exclude contributions from excitations from core to the occupied valence shell which are forbidden by the Pauli's exclusion principle. Hence half of this contribution has to be subtracted in the case of ions.
Likewise for atoms,  twice of this contribution has to be excluded from the core polarizability contribution due to fully filled valence $ns$ for Zn ($n$=4), Cd ($n$=4),  Hg ($n$=5) and $np$ for Pb ($n$=6) orbitals.
These contributions are referred as the valence-core ($\alpha_{vc}$) in our calculations.
One can calculate the static values of polarizability by substituting $\omega= 0$ in Eqs.(~\ref{Eq.9} -~\ref{Eq.11}).
	 
	\subsection{Matrix elements} 
	 In order to calculate polarizability of the monovalent ions and divalent atoms, reliable values of the matrix elements have to be calculated. In the present work, wave functions for ions and atoms are calculated using different relativistic methods. For ions, we consider relativistic all order  method confined to the single and double excitation (SD) approximation ~\cite{safronova2008all, blundell1991relativistic}. The exact wave function of the state with the closed core and single valence electron $v$ is represented as
	\begin{eqnarray}	
& &|\psi_v\rangle_{SD} = \left[1+ \sum_{ma}\rho_{ma}a_m^\dagger a_a +\frac{1}{2}\sum_{mlab}\rho_{mlab} a_m^\dagger a_l^\dagger a_b a_a\right.
\nonumber \\
& & \left.+ \sum_{m\neq v} \rho_{mv} a_m^\dagger a_v + \sum_{mla} \rho_{mlva}a_m^\dagger a_l^\dagger a_a a_v\right]|\phi_v\rangle.
						\nonumber\\
						& &	   
	 \end{eqnarray}
Here $|\phi_v\rangle$, is the mean field wavefunction constructed as $ |\phi_v\rangle = a_v^\dagger|0_c\rangle$  with $|0_c\rangle$ representing the DHF wave function of closed core and $a^\dagger$, $a$ represents creation and annihilation operators respectively whereas excitation coefficients are denoted by $\rho$. $\rho_{ma}$, $\rho_{mv}$, $\rho_{mlab}$ and $\rho_{mlva}$ being the single core, single valence, double core and double valence excitation coefficients respectively. To obtain the DHF wave functions and matrix elements for each transition, we use a set of 50 B-splines of order k = 11 for each angular momentum. The basis set orbitals are constrained to a large spherical cavity of a radius R = 220 a.u.

The required wavefunctions for divalent systems are obtained from GRASP2K code which uses MCDF approach~\cite{jonsson2013new}. In MCDF, the atomic state wavefunction (ASF) in their initial/final state can be written as the linear combination of several configurational state functions (CSFs), having the same parity and
total angular momentum, e.g.,
	\begin{equation}\label{Eq.13}
	|\psi_v\rangle_{MCDF} = \sum_{x=1}^N a_x|\phi_x\rangle,
	\end{equation}
where $x$ refers to the number of CSFs and $a_x$ is the mixing coefficient. It is important to mention that the calculation of ASFs is done by including Breit and quantum electrodynamic corrections. In order to increase the accuracy of the ASF, we consider the maximum number of CSFs in the linear contribution and, finally, retain only those which have the value of mixing coefficient greater than $10^{-3}$ . This method was used for divalent alkaline earth atoms in Ref.~\cite{shukla2020two}.

After obtaining wave functions for the aforementioned
ions and atoms, we determine the dipole-allowed (E1) matrix
element for a transition. The E1 matrix elements between the states $|\psi_v\rangle$ and $|\psi_k\rangle$ is evaluated using the following expression~\cite{PhysRevA.40.2233}
	\begin{eqnarray}\label{14}
	D_{vk} = \frac{\langle\psi_v|D|\psi_k\rangle}{\sqrt{\langle\psi_v|\psi_v\rangle \langle\psi_k|\psi_k\rangle}},
	\end{eqnarray}
	 
For practical purposes, we calculate the E1 matrix elements of some low-lying transitions, which contribute dominantly to Main term of the valence contribution using the above described method. 
Tail contribution from high lying transitions calculated using DHF method is given for ions only.
Due to some computational constraints and the sake of simplicity, the Tail contribution in the case of atoms has been neglected.

\section{Results and Discussion}\label{Sec III}
	\subsection{Dipole polarizabilities at imaginary frequencies}
	\subsubsection{Static dipole polarizability of ions}
	In Table~\ref{table1}, we present the static dipole  polarizabilities values of Zn$^+$, Cd$^+$, Hg$^+$, and Pb$^+$ heavy ions. Using Eq.~\ref{Eq.9} and ~\ref{Eq.10}, the Main and Tail terms of the valence contribution of polarizability are computed at zero frequency and given explicitly in Table~\ref{table1}. We provide the breakdown of polarizability values from every dominant transition required for the calculation of the Main term of valence contribution. Values of E1 matrix elements included in the Main term of Zn$^+$, Cd$^+$, and Hg$^+$  have been calculated in the present work while E1 matrix elements for Pb$^+$ ion have been taken from Ref.~\cite{safronova2005excitation, sahoo2005electric} which were calculated by same method as ours. The core contribution is also tabulated in the same table which has been evaluated using Eq.~\ref{Eq.11}. While $\alpha_{vc}$ contribution for Zn$^+$, Cd$^+$, and Hg$^+$ ions is almost negligible, it is notable for Pb$^+$ and affects the total polarizability value. Similar case is observed for Tail term in which significant value is observed for Pb$^+$.  
	
	In the same table, the static polarizability values of the ions are compared with the experimental and other theoretical values to enmark the validity of our values using the considered method. The polarizability values of Zn$^+$, Cd$^+$ and Hg$^+$ ions match well with the values calculated by coupled-cluster single double with triple excitations (CCSD(T)) method by Ilia\v{s} \textit{et al.} ~\cite{iliavs1999ionization}. Our polarizability value of Zn$^+$ ion deviates from the experimental value by 16\%, but it is in close agreement with other theoretical works. In recent work, Li \textit{et al.} calculated the ground state polarizability for Cd$^+$  using the DHF approximation, third-order many-body theory, and singles and doubles approximated coupled-cluster method~\cite{li2018relativistic}. The only difference in the value calculated by us and Ref.~\cite{li2018relativistic} is that they have included the  4$d^9$5$s$5$p$ configurations. The static polarizability value of Pb$^+$ in Ref.~\cite{gould2016c} was calculated using time dependent DFT (TDDFT) without including relativistic effects and fixed core approximation. {The incorporation of relativistic effects for heavy elements is required for accurate polarizability values~\cite{iliavs1999ionization}, thus the values obtained in the present work using all order method are expected to be closer to the actual values.} In a number of other studies, the present method has provided accurate values of dipole polarizability for other monovalent atoms and ions~\cite{arora2007magic} hence we can say that static dipole polarizability value of Pb$^+$ ion is also legitimate if calculated by all-order SD method. Unfortunately, we did not find any experimental measurements for static polarizability values of Cd$^+$, Hg$^+$ and Pb$^+$ ion with which we can compare our calculated results.
  
  \subsubsection{Static dipole polarizability of atoms}
	  In Table~\ref{table2}, we give the static polarizability results for Zn, Cd, Hg and Pb atoms. The breakdown of contribution of Main term from  each transition is tabulated in the same table. E1 matrix elements have been obtained from the GRASP2K code required for calculation of Main term of considered atoms. Core contribution for considered atoms is same as that for respective ions. The focus is given on calculation of valence-core $\alpha_{vc}$ correlation for atoms. Since the excitations from core to the occupied valence shell which is completely filled in case of considered atoms are not allowed, exactly twice $\alpha_{vc}$ contribution calculated in the case of ions with one valence electron has been excluded in the case of atoms. We have not included the Tail term in case of atoms. However, we anticipate very small Tail value from considered atoms except for Pb.
	  	  
	 In the same table, we also present a comparison of total value of static polarizability of the atoms calculated by us with experimental and other theoretical works. The static dipole polarizability values for Zn and Hg atoms given by Ye \textit{et al.}~\cite{ye2008dipole} using configuration interaction with a semiempirical core-polarization model potential method is found to be slightly larger than those calculated by us. Our polarizability values for Zn and Hg differ from experimental value by $\sim$9\% and $\sim$15\% respectively. Static polarizability values of Cd atom calculated by us agree well with experimental results. For Pb, the previous theoretical results has given underestimated values as compared to experimental values~\cite{pershina2008prediction, gould2016c}. The recent experimental value of Pb atom is 56 a.u. with an uncertainty of about $\pm18.2$ a.u. which is within uncertainty limits when compared to our result. However, we propose to include more transitions for more accurate polarizability for this atom.
	  
	  \begin{table*}
\caption{\label{table1} State polarizability along with contributions from various E1 reduced matrix elements to the static polarizabilities {(a.u.)} of ground state of Zn$^+$, Cd$^+$, Hg$^+$ and Pb$^+$. Main, Tail, core and valence-core contributions are given as well.  The numbers in square brackets for contribution from each transition in Main term represent powers of 10. The final results are compared with the previously estimated and available experimental values.}
	\begin{center}
\begin{tabular}{|p{1.9cm}p{1.2cm}p{1.2cm}|p{1.9cm}p{1.2cm}p{1.2cm}|p{1.9cm}p{1.2cm}p{1.2cm}|p{1.9cm}p{1.1cm}p{1.2cm}|}
\hline
\hline
\multicolumn{3}{|c|}{Zn+}  & \multicolumn{3}{c|}{Cd+} & \multicolumn{3}{c|}{Hg+} & \multicolumn{3}{c|}{Pb+}\\
& & & & & & & & & & & \\
 Transition         &  E1  &  $\alpha(0)$ & Transition & E1 & $\alpha(0)$ & Transition         & E1  &  $\alpha(0)$ & Transition & E1 & $\alpha(0)$\\
 \hline
 & & & & & & & & & & & \\	  
 $4S_{1/2}-4P_{1/2}$ & 0.189[1] & 0.537[1] & $5S_{1/2}-5P_{1/2}$ & 0.194[1] & 0.623[1] & $6S_{1/2}-6P_{1/2}$ & 0.166[1] & 0.391[1] & $6P_{1/2}-7S_{1/2}$ & 0.101[1] & 0.125[1] \\[0.5ex]
 $4S_{1/2}-4P_{3/2}$ & 0.267[1] & 0.106[2] & $5S_{1/2}-5P_{3/2}$ & 0.275[1] & 0.119[2] & $6S_{1/2}-6P_{3/2}$ & 0.235[1] & 0.666[1] & $6P_{1/2}-8S_{1/2}$ & 0.371[0] & 0.113[0] \\[0.5ex] 
 $4S_{1/2}-5P_{1/2}$ & 0.80[-1] & 0.46[-2] & $5S_{1/2}-6P_{1/2}$ & 0.10[0] & 0.77[-2] & $6S_{1/2}-7P_{1/2}$ & 0.535[0] & 0.19[0] & $6P_{1/2}-6D_{3/2}$ & 0.207[1] & 0.448[1] \\[0.5ex]
 $4S_{1/2}-5P_{3/2}$ & 0.85[-1] & 0.52[-2] & $5S_{1/2}-6P_{3/2}$ & 0.59[-1] & 0.27[-2] & $6S_{1/2}-7P_{3/2}$ & 0.366[0] & 0.88[-1] & & & \\
 $4S_{1/2}-6P_{1/2}$ & 0.86[-1] & 0.45[-2] & $5S_{1/2}-7P_{1/2}$ & 0.113[0] & 0.85[-2] & & & & & & \\[0.5ex]
 $4S_{1/2}-6P_{3/2}$ & 0.143[0] &  0.13[-1] & $5S_{1/2}-7P_{3/2}$ & 0.117[0] & 0.91[-2] & & & & & & \\[0.5ex]
 $\alpha_{val}^{Main}$ & & 15.98 & $\alpha_{val}^{Main}$ & & 18.14 & $\alpha_{val}^{Main}$ & & 10.84 & $\alpha_{val}^{Main}$ & & 5.84 \\[0.5ex]
 $\alpha_{val}^{Tail}$ & & 0.02 & $\alpha_{val}^{Tail}$ & & 0.01 & $\alpha_{val}^{Tail}$ & & 0.06 & $\alpha_{val}^{Tail}$ & & 2.67 \\[0.5ex]
 $\alpha_{vc}$ & & 0.006 & $\alpha_{vc}$ & & -0.02 & $\alpha_{vc}$ & & -0.04 & $\alpha_{vc}$ & & -2.28 \\[0.5ex]
 $\alpha_{c}$ & & 2.05 & $\alpha_{c}$ & & 5.28 & $\alpha_{c}$ & & 8.21 & $\alpha_{c}$ & & 16.30 \\[0.5ex]
 Total & & 18.05 & Total & & 23.41 & Total & & 19.07 & Total & & 22.52 \\[0.5ex] 
 Experiment & & 15.54~\cite{kompitsas1994rydberg} & & & & & & & & & \\[0.5ex]
 Others & & 18.84~\cite{iliavs1999ionization} & Others & & 23.68~\cite{iliavs1999ionization} & Others & & 19.36~\cite{iliavs1999ionization} & Others & & 23.5~\cite{gould2016c} \\[0.5ex]
 & & 17.90~\cite{gould2016c} & & & 23.1~\cite{gould2016c} & & & 17.50~\cite{gould2016c} & & &  \\[0.5ex]
 & & & & & 25.21~\cite{li2018relativistic} & & & & & & \\
 
 \hline
   
	 \end{tabular}	   
	 \end{center}
	 \end{table*} 

	 \begin{table*}
\caption{\label{table2}State polarizability along with contributions from various E1 reduced matrix elements to the static polarizabilities {(a.u.)} of ground state of Zn, Cd, Hg and Pb. Main, core and valence-core contributions are given as well. The final results are compared with the previously estimated and available experimental results. The numbers in square brackets for contribution from each transition in Main term represent powers of 10. The uncertainty in experimental values are given in the parentheses.}

	\begin{center}
\begin{tabular}{|p{1.9cm}p{1.2cm}p{1.2cm}|p{1.9cm}p{1.2cm}p{1.2cm}|p{1.9cm}p{1.2cm}p{1.2cm}|p{1.9cm}p{1.1cm}p{1.2cm}|}
\hline
\hline
\multicolumn{3}{|c|}{Zn}  & \multicolumn{3}{c|}{Cd} & \multicolumn{3}{c|}{Hg} & \multicolumn{3}{c|}{Pb}\\
& & & & & & & & & & &\\
 Transition          & E1 &  $\alpha(0)$ & Transition & E1  & $\alpha(0)$ & Transition  & E1 &  $\alpha(0)$ & Transition & E1  & $\alpha(0)$\\
 \hline
 & & & & & & & & & & &\\	  
 $4 {}^1S_0-4 {}^3P_1$ & 0.17[-3] & 0.748[-5] & $5 {}^1S_0-5 {}^3P_1$ & 0.208[-1] & 0.278[-3] & $6 {}^1S_0-6 {}^3P_1$ & 0.118[0] & 0.437[0] & $6 {}^3P_0-7 {}^3P_1$ & 0.16[-1] & 0.692[1] \\[0.5ex]
  $4 {}^1S_0-4 {}^1P_1$ & 0.103[2] & 0.322[2] & $5 {}^1S_0-5 {}^1P_1$ & 0.949[1] & 0.318[2] & $6 {}^1S_0-6 {}^1P_1$ & 0.725[1] & 0.196[2] & $6 {}^3P_0-7 {}^1P_1$ & 0.80[-1] & 0.213[0] \\[0.5ex] 
 $4 {}^1S_0-5 {}^3P_1$ & 0.40[-4] & 0.839[-6]& $5 {}^1S_0-6 {}^3P_1$ & 0.16[-2] & 0.407[-4] & $6 {}^1S_0-7 {}^3P_1$ & 0.93[-3] & 0.196[-4] & $6 {}^1S_0-7 {}^1P_1$ & 0.491[1] & 0.362[2] \\[0.5ex]
 $4 {}^1S_0-5 {}^1P_1$ & 0.466[0] & 0.108[1] & $5 {}^1S_0-6 {}^1P_1$ & 0.488[1] & 0.119[2] & $6 {}^1S_0-7 {}^1P_1$ & 0.554[-1] & 0.114[0] & $6 {}^3P_0-6 {}^3D_1$ & 0.229[1] & 0.704[1] \\[0.5ex]
 $4 {}^1S_0-6 {}^3P1$ & 0.40[-4] & 0.899[-6] & $5 {}^1S_0-7 {}^3P_1$ & 0.4[-4] & 0.865[-6] & $6 {}^1S_0-8 {}^3P_1$ & 0.143[0] & 0.273[0] & & & \\[0.5ex]
 $4 {}^1S_0-6 {}^1P_1$ & 0.18[-1] & 0.251[-3] & $5 {}^1S_0-7 {}^1P_1$ & 0.282[0] & 0.628[0] & $6 {}^1S_0-8 {}^1P_1$ & 0.428[-1] &  0.815[-3] & & &\\[0.5ex]
 $\alpha_{val}^{Main}$ & & 33.29 & $\alpha_{val}^{Main}$ & & 44.36 & $\alpha_{val}^{Main}$ & & 20.51 & $\alpha_{val}^{Main}$ & & 50.39 \\[0.5ex]
 $\alpha_{vc}$ & & -0.0013 & $\alpha_{vc}$ & & -0.04 & $\alpha_{vc}$ & & -0.08 & $\alpha_{vc}$ & & -4.58 \\[0.5ex]
 $\alpha_{c}$ & & 2.05 & $\alpha_{c}$ & & 5.28 & $\alpha_{c}$ & & 8.21 & $\alpha_{c}$ & & 16.30 \\[0.5ex]
 Total & & 35.33 & Total & & 49.61 & Total & & 28.65 & Total & & 61.90 \\[0.5ex] 
 Experiment & \multicolumn{2}{r|}{38.80(0.3)} & Experiment & \multicolumn{2}{r|}{49.65(1.46)} & Experiment & \multicolumn{2}{r|}{33.91(0.34)} & Experiment & \multicolumn{2}{r|}{56.0(18.2)} \\[0.5ex]
 & & ~\cite{goebel1996theoretical} & & &  ~\cite{goebel1995dispersion} & & & ~\cite{singh2015rigorous} & & & ~\cite{ma2015measured} \\[0.5ex]
 Others & \multicolumn{2}{r|}{37.6~\cite{kello1995polarized}} & Others & \multicolumn{2}{r|}{46.8~\cite{kello1995polarized}} & Others & \multicolumn{2}{r|}{31.2~\cite{kello1995polarized}} & Others & \multicolumn{2}{r|}{46.96~\cite{pershina2008prediction}} \\[0.5ex]
 & \multicolumn{2}{r|}{38.4~\cite{gould2016c}} & & \multicolumn{2}{r|}{46.7~\cite{gould2016c}} & & \multicolumn{2}{r|}{33.5~\cite{gould2016c}} & & \multicolumn{2}{r|}{47.9~\cite{gould2016c}}\\[0.5ex]
 & \multicolumn{2}{r|}{38.12~\cite{ye2008dipole}} & & \multicolumn{2}{r|}{44.63~\cite{ye2008dipole}} & & \multicolumn{2}{r|}{31.32~\cite{ye2008dipole}} & & \multicolumn{2}{r|}{ }\\[0.5ex]
 
 \hline
	 \end{tabular}   
	 \end{center}
	 \end{table*}
	 
	 \subsubsection{Dynamic dipole polarizability at imaginary frequency}
\begin{figure}
	  \includegraphics[width=\columnwidth,keepaspectratio]{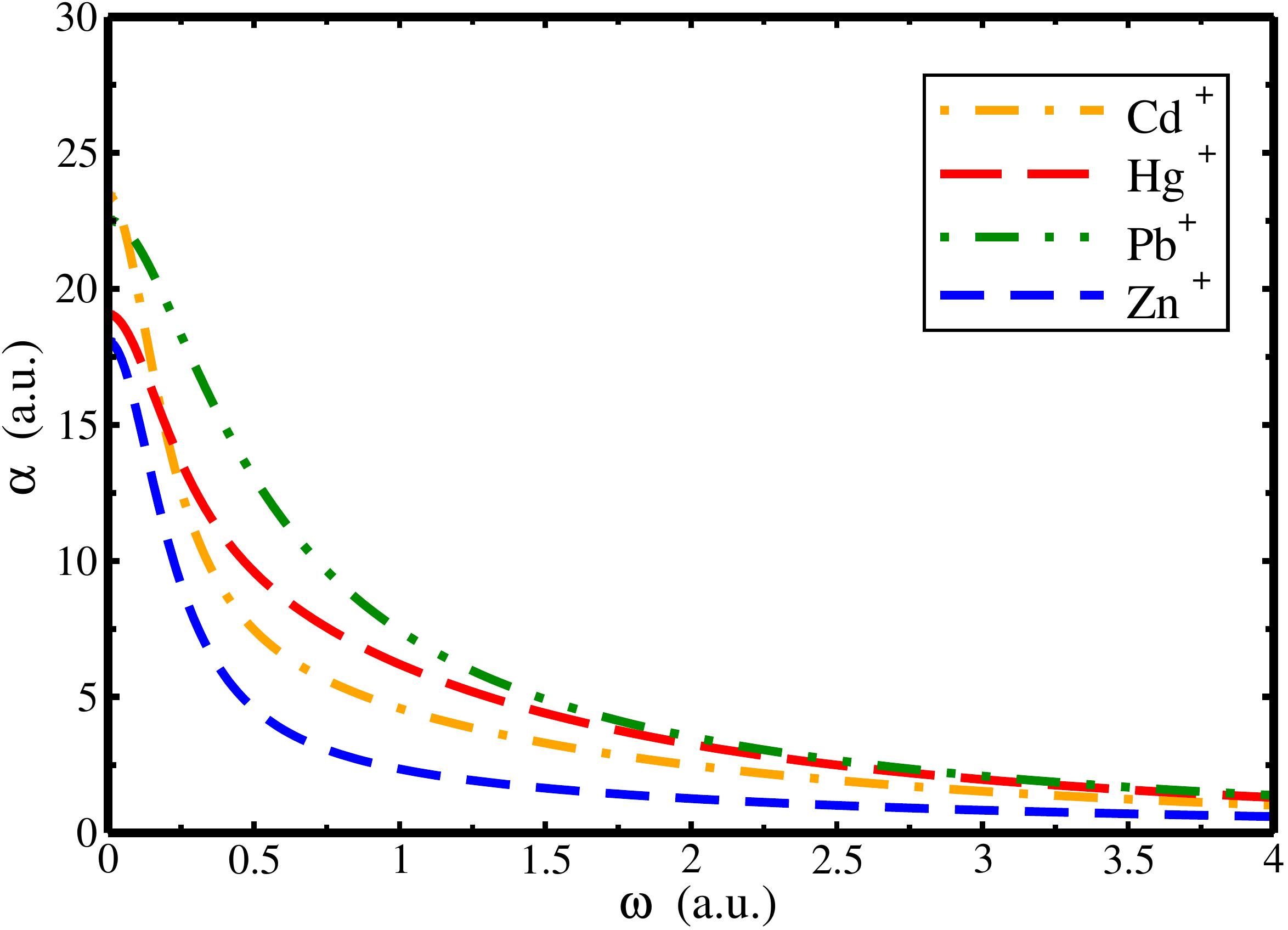}
	  \caption{\textcolor{red}{(Color online) Dynamic dipole polarizability $\alpha$ (a.u.) at imaginary frequencies of Zn$^+$ (blue dashed curve), Cd$^+$ (orange dotted dashed curve), Hg$^+$ (red long dashed curve) and Pb$^+$ (green double dotted dashed curve).}}
	  \label{Fig.1}
	  \end{figure}

\begin{figure}
	  \centering
	  \includegraphics[width=\columnwidth,keepaspectratio]{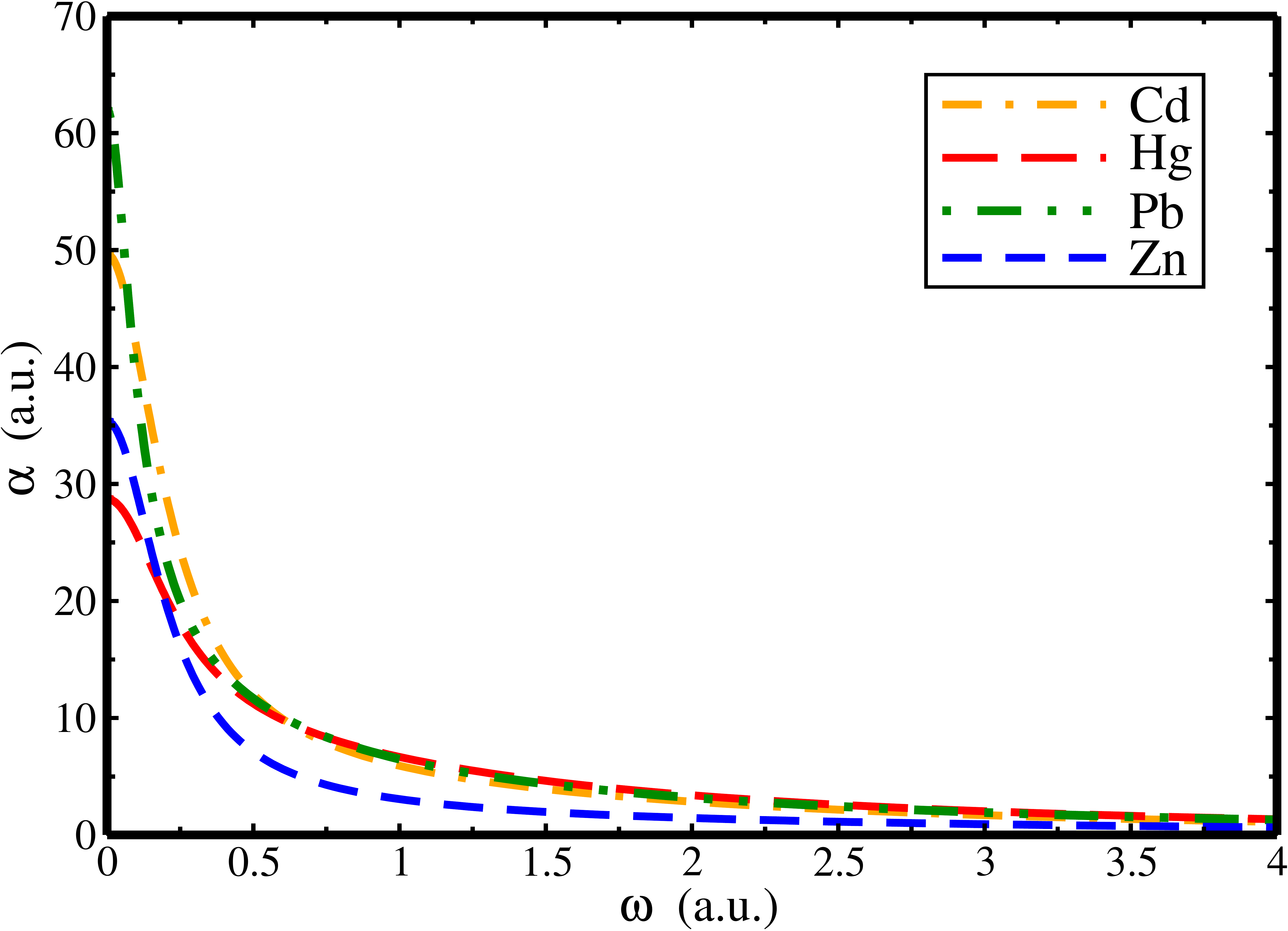}
	  \caption{\textcolor{red}{(Color Online) Dynamic dipole polarizabilities $\alpha$ (a.u.) at imaginary frequencies of Zn (blue dashed curve), Cd (orange dotted dashed curve), Hg (red long dashed curve) and Pb (green double dotted dashed curve)}.}
	  \label{Fig.2}
	  \end{figure}

	  We determine the dynamic polarizability values at different frequencies using the same method which has been used for evaluation of static polarizability and expect our values to be reliable. Dynamic dipole polarizabilities of ions and atoms at imaginary frequencies are presented in Fig.~\ref{Fig.1} and Fig.~\ref{Fig.2} respectively. Tabulated values of $\alpha(\iota\omega)$ for considered ions and atoms are given in Supplementary materials (SM)~\cite{SM}. With increase of frequency, polarizability decreases and reaches a small value. This trend is seen for both ions and atoms.  From Fig.~\ref{Fig.1}, one notices that at short frequency values, the dipole polarizabilities of Cd$^+$ and Pb$^+$ ions are comparable but with an increase in  frequency, the polarizability  of Cd$^+$ decreases more rapidly as compared to Pb$^+$ ion. Dynamic polarizability of Cd$^+$ is even lower than polarizability of Hg$^+$ for $\omega > 0.25$ a.u. For Zn$^+$ ion, the polarizability remains lowest throughout the frequency regime as compared to other ions. Similarly as shown in Fig.~\ref{Fig.2}, for atoms the static polarizability of Zn is larger than Hg at short $\omega$ but decreases rapidly for Zn  as compared to Hg as $\omega$  increases.  For $\omega > 1$ a.u., the polarizability values of Hg atom are more as compared to value of Zn and Cd and get closer to the polarizability of Pb atom. If we compare polarizability values among atoms and ions, the lowest value throughout the considered frequency range is observed for Zn$^+$ ion whereas the largest values for Pb and Cd atoms depending upon the frequency value. These values have been used for computing $C_3$ dispersion coefficients as a function of separation distance as discussed in the next section. 
	  
\subsection{$C_3$ Coefficients for graphene}
\begin{figure}	
	 \centering  \includegraphics[width=\columnwidth,keepaspectratio]{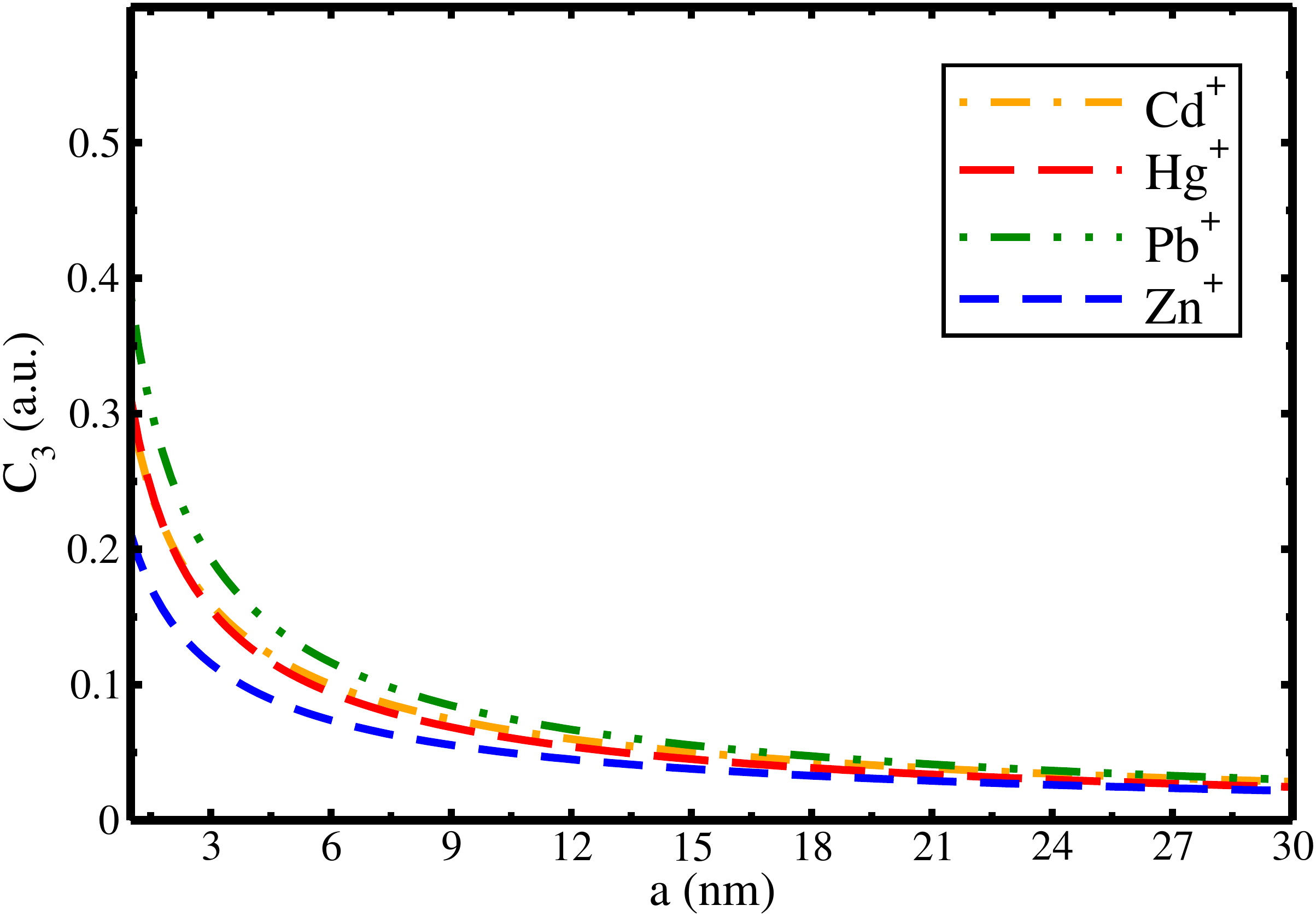}
	  \caption{\textcolor{red}{(Color online) $C_3$ dispersion coefficients (a.u.) for interaction of Zn$^+$ (blue dashed curve), Cd$^+$ (orange dotted dashed curve), Hg$^+$ (red long dashed curve) and Pb$^+$ (green double dotted dashed curve) with graphene layer as a function of separation distance $a$ (nm).}}
	  \label{Fig.3}
	\end{figure}

\begin{figure}	
	 \centering  \includegraphics[width=\columnwidth,keepaspectratio]{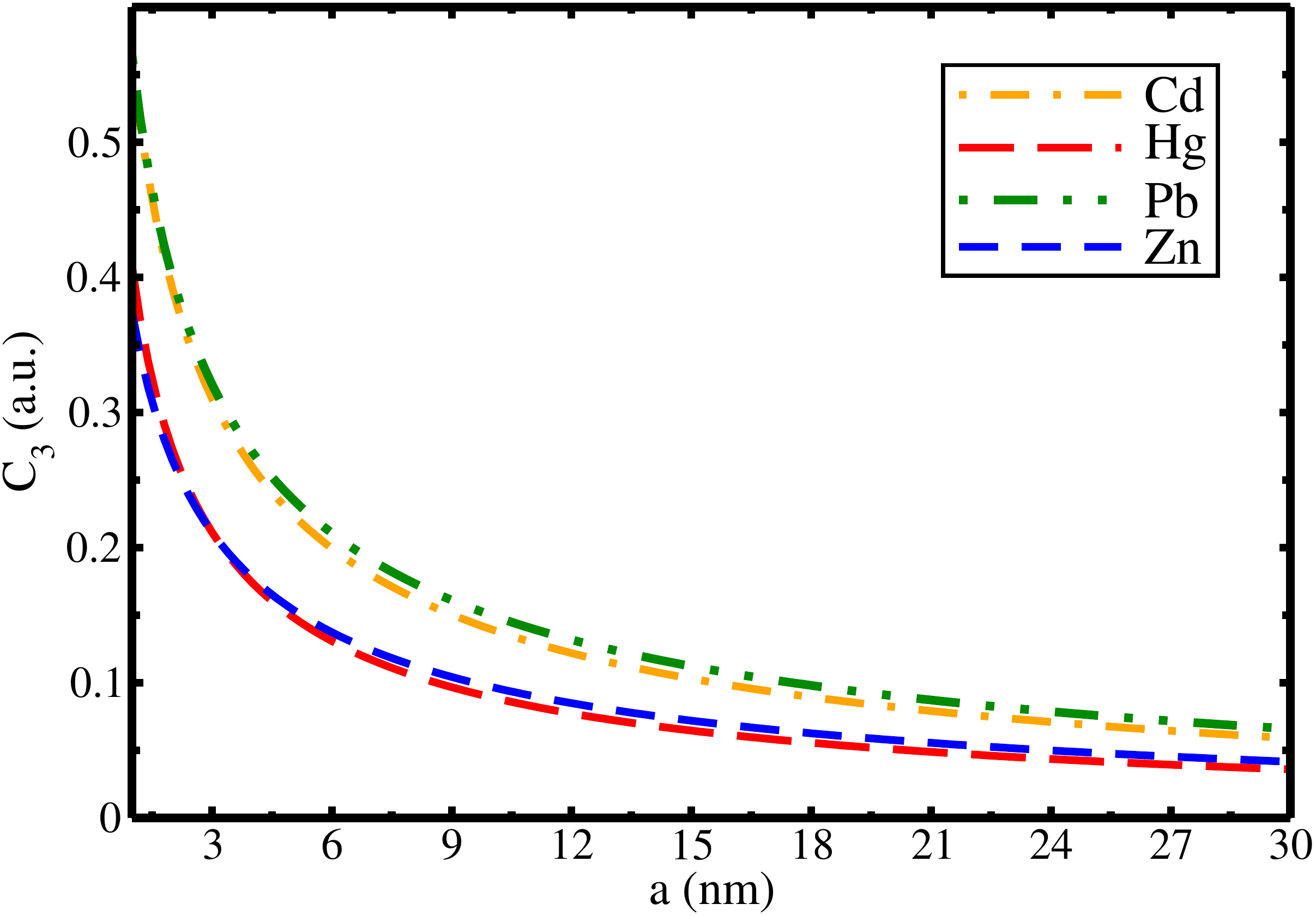}
	  \caption{\textcolor{red}{(Color online) $C_3$ dispersion coefficients (a.u.) for interaction of Zn (blue dashed curve), Cd (orange dotted dashed curve), Hg (red long dashed curve) and Pb (green double dotted dashed curve) atoms with graphene layer as a function  of separation distance $a$ (nm).}}
	  \label{Fig.4}
	\end{figure}

	  In this section, we present the dispersion coefficients between graphene layer with $\Delta = $ 0.01 eV and heavy elements as a function of separation distance. As shown in Fig.~\ref{Fig.3} and Fig.~\ref{Fig.4}, $C_3$ coefficients exhibit an inverted yield curve i.e., $C_3$ values decrease with increase in distance. This nature of the curve is perceived for every element. Coefficients reach a value less than 0.1 a.u. for distance greater than 30 nm. Among the ions, the largest $C_3$ value is observed for Pb$^+$ indicating stronger interaction with graphene layer whereas Zn$^+$ is least attracted as shown in Fig.~\ref{Fig.3}. The respective $C_3$ values for Cd$^+$ and Hg$^+$ ions are approximately the same. In the case of atoms, the large $C_3$ values have been observed for Pb and Cd. At $a$ = 1 nm, $C_3$ values are 0.564 and 0.561 for Pb and Cd atom respectively. However, for  separation distance $a > 7$ nm, the difference in $C_3$ coefficients for Pb and Cd become appreciable as displayed in Fig.~\ref{Fig.4}. Zn and Hg are least attracted towards graphene. To best of our knowledge, we did not find any literature on $C_3$ coefficient values for interaction of heavy ions and atoms with carbon-based nanostructures. However, a comparison has been made based on physisorption of heavy atoms with previous DFT study~\cite{shtepliuk2017interaction}. For atoms, our results are in accordance with the DFT study for physisorption of heavy atoms on graphene layer~\cite{shtepliuk2017interaction}. In DFT study, vdW interactions were described between the adsorbed atom and graphene and the strength of interactions were analysed as a function of binding energy and charge transfer. The sequence of reducing binding energy for atoms was reported as Pb $>$ Cd $>$ Hg~\cite{shtepliuk2017interaction}. A similar trend is observed in our study where we have analysed the strength of interaction on the basis of $C_3$ values. Shtepliuk \textit{et al.}~\cite{shtepliuk2017interaction} also studied the interaction of ions with graphene which resulted in chemisorption. Since our study provides the result for physisorption of microparticles on the material wall, hence we do not make a similar comparison for ions with Ref.~\cite{shtepliuk2017interaction}. It is important to note that these ions with large nuclear charge Z, which we have considered in the present work are only singly charged, (\textit{i.e.,} having residual unity charge), thus the effect of the latter may not be too significant as compared to the coulomb potential of heavier Z atoms in the calculation of matrix elements with which we have evaluated the $C_3$ coefficients. The overall charge present on the ions leads to stronger Coulomb interactions as compared to weak vdW attractions with both graphene and CNT wall.  The considered theory only provides the information regarding the weak vdW forces between the microparticle and considered substrates,  which on comparison reveals that  the selectivity and sensitivity of graphene and CNT to adsorb heavy atoms  is more as compared to ions.     
	
	\subsection{$C_3$ Coefficients for CNT}
\begin{figure}	
	 \centering  \includegraphics[width=\columnwidth,keepaspectratio]{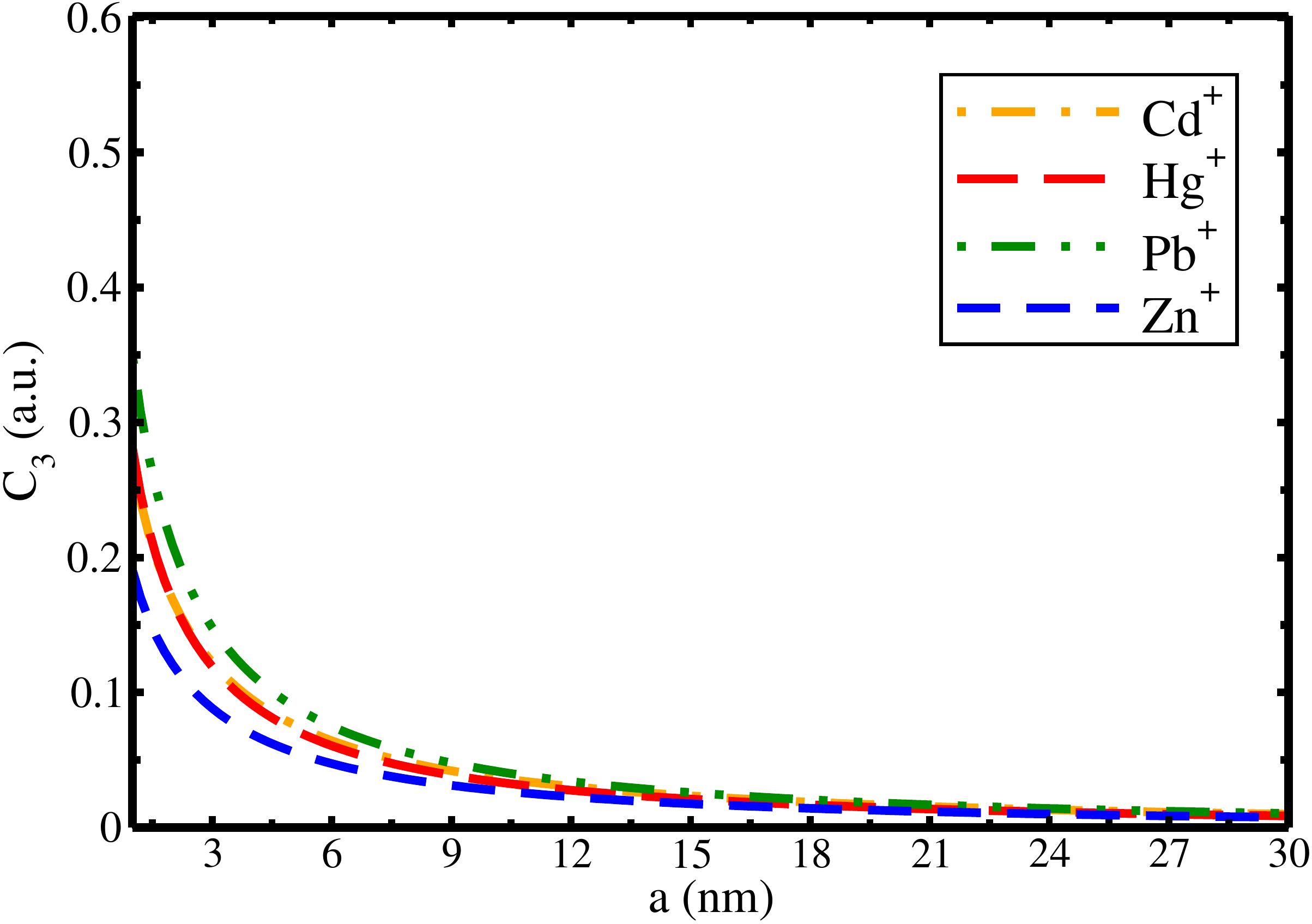}
	  \caption{\textcolor{red}{(Color online) $C_3$ dispersion coefficients (a.u.) for interaction between Zn$^+$ (blue dashed curve), Cd$^+$ (orange dotted dashed curve), Hg$^+$ (red long dashed curve) and Pb$^+$ (green double dotted dashed curve) with CNT of radius $R$ = 6 nm as a function of separation distance $a$ (nm).}}
	  \label{Fig.5}
	\end{figure}

\begin{figure}	
	 \centering  \includegraphics[width=\columnwidth,keepaspectratio]{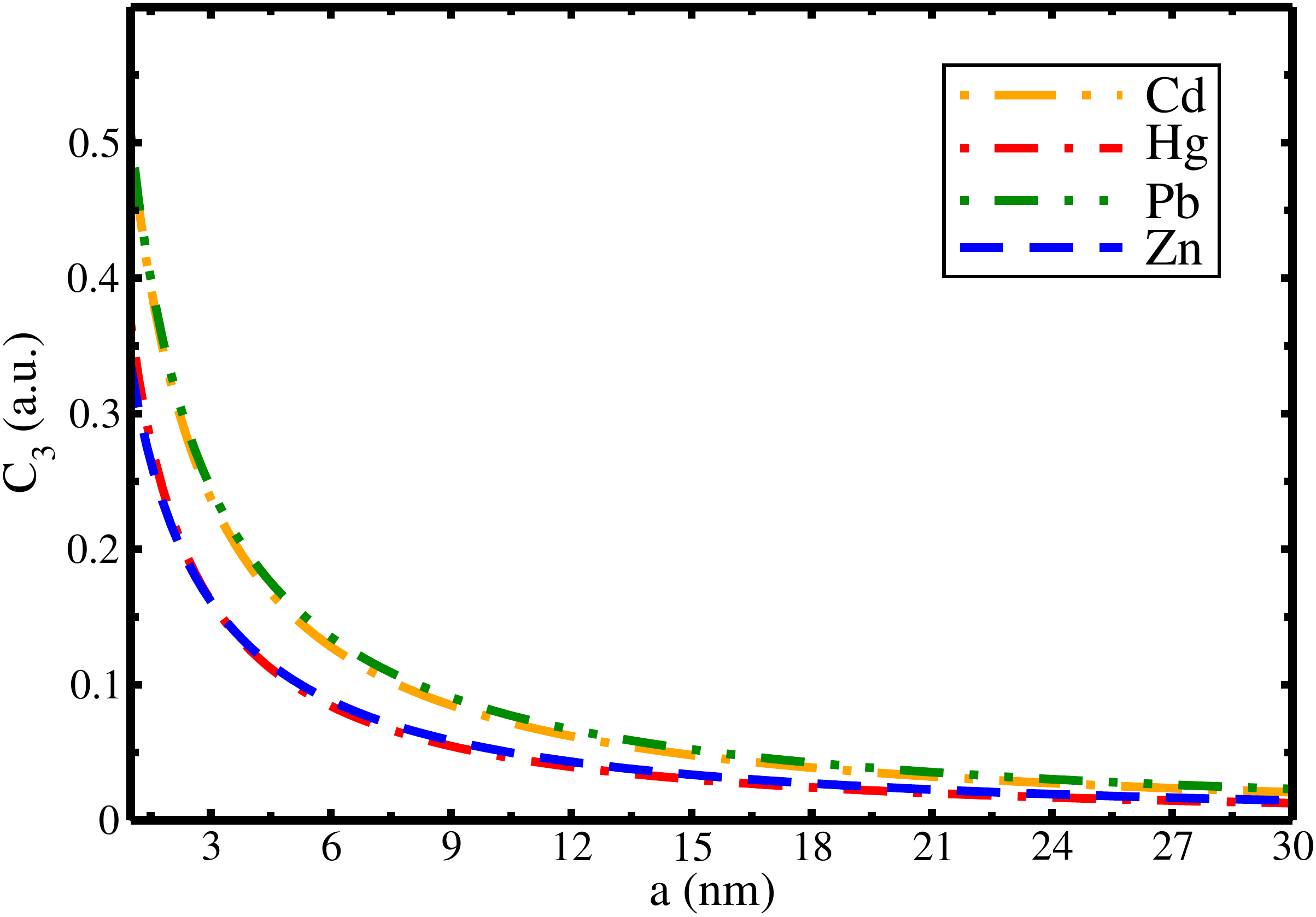}
	  \caption{\textcolor{red}{(Color online) $C_3$ dispersion coefficients (a.u.) for interaction between Zn (blue dashed curve), Cd (orange dotted dashed curve), Hg (red long dashed curve) and Pb (green double dotted dashed curve) atom with CNT of radius $R$ = 6 nm as a function of separation distance $a$ (nm).}}
	  \label{Fig.6}
	\end{figure}

In addition to graphene, we present the dispersion coefficient for CNT. Fig.~\ref{Fig.5} and Fig.~\ref{Fig.6} represent the influence of separation distance on dispersion coefficients evaluated for CNT of radius of 6 nm with heavy ions and atoms respectively. The $C_3$ coefficients and hence interaction are dominant at smaller distances for all the elements. Similar to the case of graphene, the interaction is strongest for Pb$^+$ and weakest for Zn$^+$ with CNT.  For atoms, CNT offers a stronger potential to Pb and Cd atoms and weak potential to Zn atom. 

\begin{figure}	
	 \centering  \includegraphics[width=\columnwidth,keepaspectratio]{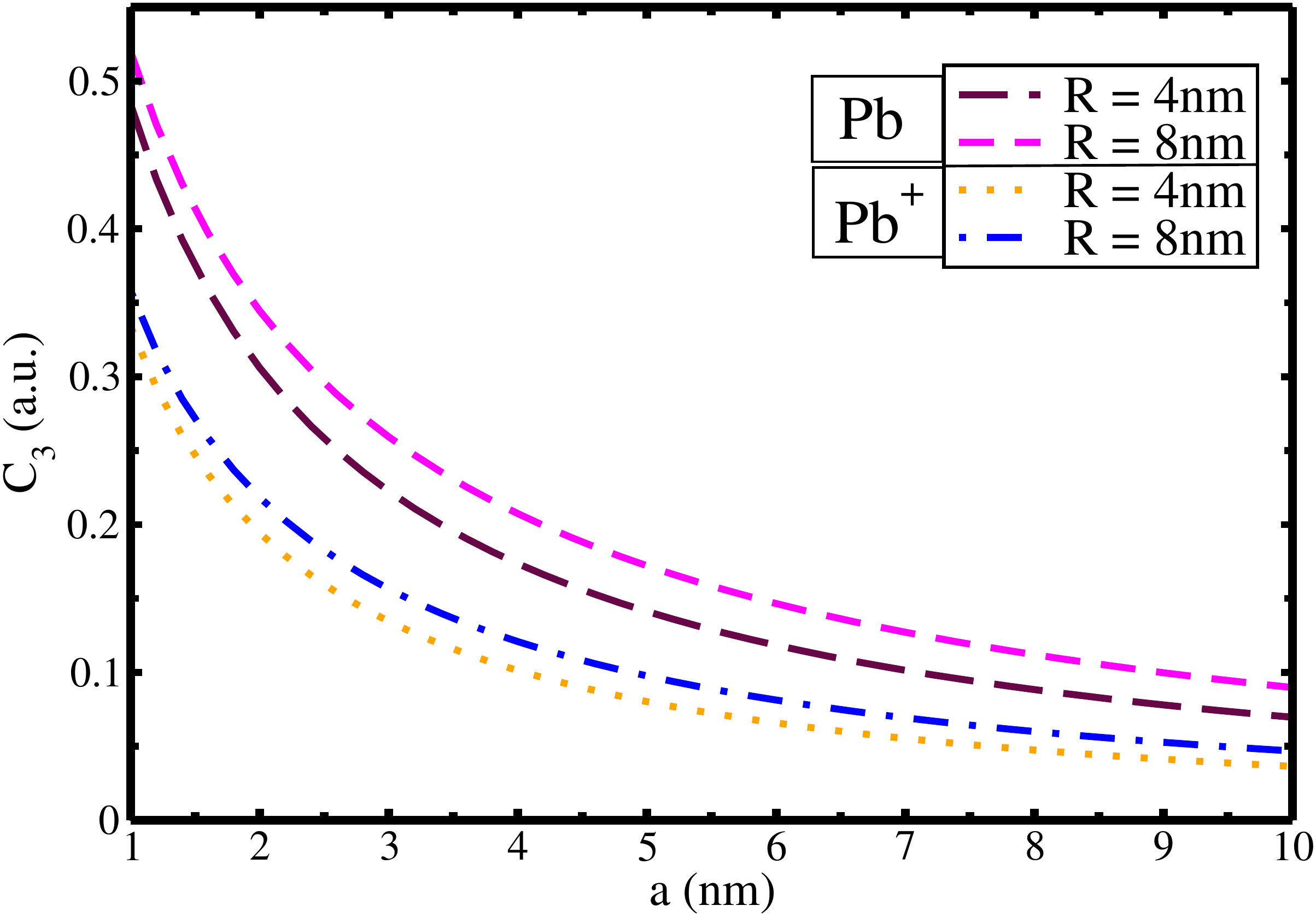}
	  \caption{\textcolor{red}{(Color online) $C_3$ dispersion coefficients (a.u.) for Pb atom (purple long dashed curve for $R$ = 4 nm, pink dashed curve for $R$ = 8 nm) and Pb$^+$ ion (orange dotted curve for $R$ = 4 nm, blue dotted dashed curve for $R$ = 8 nm) for two different values of radius of CNT.}}
	  \label{Fig.7}
	\end{figure}

The radius of CNT has been an important parameter in hydrogen storage applications. It has been known that larger radius of CNT imparts more gravimetric storage amount for hydrogen storage~\cite{weng2007atomistic}. This motivated us to study the effect of radius of CNT on $C_3$ coefficients. Fig. \ref{Fig.7} demonstrates the effect of radius of CNT on dispersion coefficients. CNT with larger radius i.e., 8 nm has more potential to adsorb Pb species. This can be attributed due to greater exposure of carbon atoms towards ions and atoms with increase in radius of CNTs.   
	
	\subsection{Dispersion coefficient for different gap parameter}
\begin{figure}	
	 \centering  \includegraphics[width=\columnwidth,keepaspectratio]{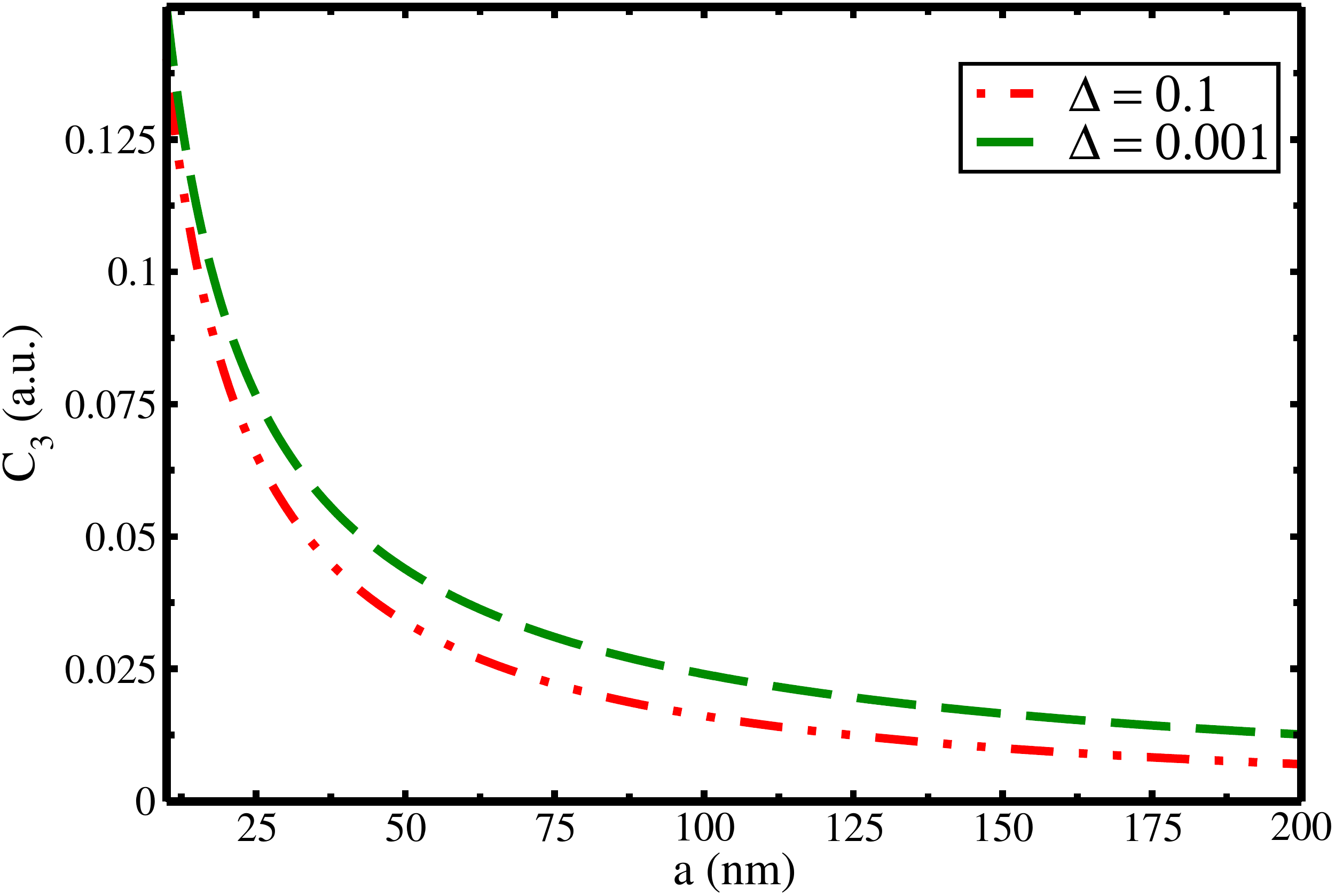}
	  \caption{\textcolor{red}{(Color online) $C_3$ dispersion coefficients (a.u.) for Pb atom with graphene wall as a function of separation distance $a$ (nm) for two different values of gap parameter $\Delta$ (green long dashed curve for $\Delta$ = 0.1 and red double dotted dashed curve for $\Delta$ = 0.001) in eV.}}
	  \label{Fig.8}
	\end{figure}

	Most of the previous studies on dispersion coefficients for graphene and CNT have taken the upper bound of the gap parameter $\Delta$ as 0.1 eV~\cite{churkin2011dispersion, chaichian2012thermal, arora2014coefficients}. The effect of this parameter was shown while studying the interactions between alkali atoms and graphene layer in our previous study~\cite{kaur2014emending}. In the present work, we investigate the dependency of gap parameter on $C_3$ coefficients as well.  Since the interaction of Pb atom is most prominent among all the elements considered in this study, we present the effect of gap parameter on dispersion coefficient between graphene layer and Pb atom. To find the influence of $\Delta$, we choose two different values of  $\Delta$ as 0.1 eV and 0.001. eV The dispersion coefficient is seen to increase by only 0.41 \% at separation distance of $a = 1$ nm with decrease of $\Delta$ from 0.1 eV to 0.001 eV. When investigated the same at a larger separation distance of 100 nm and 200 nm, the percentage increase in the $C_3$ coefficient is find to be about 49 and 80 respectively. This is presented in Fig.~\ref{Fig.8} where gap between the two curves of different $\Delta$ increases with increase in separation.
	
	\subsection{Comparison of graphene and CNT}
	Since we studied the dispersion interactions for two carbon-based materials, it is important to compare these two and prognosticate a better material for physisorption of these heavy elements. Both the materials show similar trend for adsorption of ions and atoms and are selective towards Pb and Cd atoms. The weakest interaction is observed for Zn$^+$ ion. 
However, when a comparison is drawn for interaction of a microparticle with graphene and CNT it was found that graphene provided larger $C_3$ coefficient value and hence stronger interaction as compared to CNT.
Contrarily,  a number of studies can be found in literature where CNT is widely accepted for physisorption applications as compared to graphene. The reason behind this is ability of experimentalists to tailor the properties and structure of CNTs with ease whereas the bulk preparation of pristine graphene is a major bottleneck that needs a direction~\cite{petit1999tuning}.
	
	\section{Conclusion}\label{Sec IV}
 To conclude, we have probed the dispersion coefficients for Zn$^+$, Cd$^+$, Hg$^+$, Pb$^+$, Zn, Cd, Hg and Pb with graphene and CNT walls. We have provided the dynamic dipole polarizability values for both heavy ions and atoms using sum-over-states approach. The interactions between heavy elements and material wall as a result of dispersion $C_3$ coefficients has been found to be maximum for Pb atom and ion at short separations. The result of interaction studies by our methodology is in agreement with interactions studied by DFT for heavy atoms. CNT also shows the potential for interaction of heavy elements following similar trend as that of graphene. We also deduce that graphene is more sensitive for interaction of the considered elements as compared to CNT. The obtained results could be useful for the formation of highly sensitive and selective sensors for detection of heavy ions and atoms.
 
 \section{Acknowledgements}
The authors, B. A. is thank-full to the SERB-TARE(TAR/2020/000189), New Delhi, India  for research grant. While R. S. is thank-full for the sanction of the research grant no: CRG/2020/005597 by SERB-DST, New Delhi, India.     

\end{document}